\documentclass[12pt]{article}
\usepackage[centertags]{amsmath}
\newcommand{\be}{\begin{equation}}
\newcommand{\ee}{\end{equation}}
\usepackage[centertags]{amsmath}
\usepackage{amsfonts}
\newcommand{\nn}{\noindent}
\begin{document}
%%%%%%%%%%%%%%%%%%%%%%%%%%%%%%%%%%%%%%%%%%%%%%%%%%%%%%%%%%%%%%%%%%%
\title{\bf Bogomolny equations for vortices in the noncommutative torus}
\author{P. Forg\'acs$^a$, G.S.~Lozano$^b$\thanks{Associated with CONICET}\,,\\
E.F.~Moreno$^c$$^*$ \, and \,
%\thanks{Associated with CONICET}
F.A.~Schaposnik$^d$\thanks{Associated with CICBA}
\\
{\normalsize\it $^a$Laboratoire de Math\'ematiques et Physique Th\'eorique}\\
{\normalsize\it CNRS/UMR 6083, Universit\'e de Tours}\\
{\normalsize\it Parc de Grandmont, 37200 Tours, France}
\\
{\normalsize\it $^b$ Departamento de F\'\i sica, FCEyN,
Universidad de Buenos Aires}\\ {\normalsize\it Pab.1, Ciudad
Universitaria, Buenos Aires,Argentina}
\\
{\normalsize\it $^c$Department of Physics,West Virginia University}\\
{\normalsize\it Morgantown, West Virginia 26506-6315, U.S.A.}\\
 {\normalsize\it $^d$Departamento de F\'\i
sica, Universidad Nacional de La Plata}\\ {\normalsize\it C.C. 67,
1900 La Plata, Argentina}}

\maketitle
%===================================================================
%===================================================================
\begin{abstract}
We derive Bogomolny-type equations for the Abelian Higgs model
defined on the noncommutative torus and discuss its vortex like
solutions. To this end, we carefully analyze how periodic boundary
conditions have to be handled in noncommutative space and
discussed how vortex solutions are constructed. We also consider
the extension to an $U(2)\times U(1)$ model, a simplified
prototype of the noncommutative standard model.
\end{abstract}
\section{Introduction}

Construction of noncommutative solitons and instantons has been a
field of intense activity after the revival of
 field theories in noncommutative space, in connection with string
 theory and brane dynamics (see for example \cite{DN} for
 references on this issue). Not only the noncommutative
counterparts of vortices, monopoles and other localized solutions
in ordinary space were constructed but regular stable solutions
which become singular in the commutative limit were also
discovered (see for example \cite{FS} for a complete list of
references). Concerning static classical solutions of the Abelian
Higgs model in the noncommutative plane, both BPS and non BPS
vortices have been constructed and its moduli space studied in
detail \cite{JMW}.

In the present work we consider vortex solutions in the Abelian
Higgs model defined on the noncommutative torus and then extend
the analysis to the case of a $U(2)\times U(1)$ symmetry. This is
motivated by the fact that, in commutative space, one can find
stable solutions that correspond to  periodic arrays of vortices
in theories with gauge field coupled to Higgs scalars. Moreover,
the analysis of such kind of arrays is equivalent to the study of
models defined on the torus. This fact has been exploited in the
search of vortex solutions in the Salam-Weinberg model where the
only stable solutions correspond to such type of arrays
\cite{BL1}. Hence our results can be seen as a first step along
this line in its noncommutative version.

Despite the fact that the 2 dimensional torus is one of the
simplest examples of noncommutative space, no discussion of the
BPS equations and their solution for the Maxwell-Higgs model has
been carried out. In this respect,  our work  fills in this gap
and also opens the possibility of studying non-Abelian extensions
related to the noncommutative version of the Salam-Weinberg
theory. Bogomolny equations for the Abelian Higgs model on a two
dimensional torus have been first considered by Shah and Manton
\cite{SM}. More recently, Gonzalez Arroyo and Ramos \cite{GAR}
have analyzed them in detail and presented a high precision
approximation scheme.
%First order equations for non-abelian
%vortices have been derived for the standard model in Ref.
%\cite{BL2}. We extend the analysis of Ref.\cite{BL2} for a $U(2)
%\times U(1)$ model on the non-commutative torus.

The paper is organized as follows: we introduce  in section 2 the
noncommutative torus ${\cal T}$ (and noncommutative parameter
$\theta$) and discuss periodicity conditions for gauge and matter
fields. We show that consistency of gauge transformations and
periodicity conditions naturally leads to the introduction of a
scaled torus $\bar {\cal T}$ and a $\theta$-depending scaled gauge
charge. Then, in section 3 we discuss the dynamics of the
Maxwell-Higgs model showing that the role of the scaled torus
becomes crucial in the definition of  gauge invariant expressions
for the energy and magnetic flux as well as for the obtention of
covariant BPS equations. We present a particular solution to these
equations and we also discuss the strategy to obtain general
vortex like solutions, analogous to that leading to numerical
solutions in the commutative torus \cite{GAR}. Finally, in section
4 we extend the discussion to the case of a $U(2)\times U(1)$
Lagrangian for which we also write the BPS equations and indicate
how one should look for their solution. We leave for an Appendix
the derivation of some results needed to implement periodic
boundary conditions on the noncommutative torus.

\section{ Gauge and matter fields  on the noncommutative torus}

Let us consider noncommutative $2+1$ dimensional space-time with
coordinates satisfying \be [x,y] = i \theta \, , \;\;\;\;\;\;
[x,t]=[y,t] = 0 \label{ncspace} \ee
Our model will be defined on a spatial torus ${\cal T}$ with
periods $(L_1,L_2)$.

We shall be interested in a $U(1)$ gauge theory with
 Higgs scalars $\phi$ in the
fundamental  representation coupled to
 gauge fields $A_i$. The fields transform under the $U(1)$ gauge group
 according to
\begin{align}
A_{i} \, &\to \, A^{(V)} =  V^{-1}\, A_{i} \, V + \frac{i}{g}\,
V^{-1}\,
\partial_{i} \, V \\
\Phi \, &\to \, \Phi^{(V)} = V^{-1}\, \Phi \label{conv1}
\end{align}

As in ordinary space, a scalar field on the noncommutative torus
can be defined as a function $\phi(x,y)$ which is periodic up to
gauge transformations. That is,
\begin{align}
\phi(x+L_1,y) &= U_1(x,y)\, \phi(x,y) = \phi^{U_1^{-1}}(x,y)\nonumber \\%
\phi(x,y+L_2) &= U_2(x,y)\, \phi(x,y) =  \phi^{U_2^{-1}}(x,y)\label{comm-1}
\end{align}
where $U_1$ and $U_2$ are $U(1)$ gauge transformations. Concerning
 gauge fields, boundary conditions are
\begin{align}
A_{i}(x+L_1,y) &=  A^{(U_1^{-1})}_i(x,y) \\
A_{i}(x,y + L_2) &= A_i^{(U_2^{-1})}(x,y)
 \label{bca}
\end{align}
Consistency of the precedent equations implies
\be U_{2}(x+L_1,y)\, U_1(x,y) = U_1(x,y+L_2)\, U_{2}(x,y)
\label{concon}\ee
Note that eq.(\ref{concon}) coincides with the well-known consistency condition
for the commutative torus (See for example \cite{rev} and references therein).

A particular solution to eq.(\ref{concon}) is
\be U_1(x,y) = e^{i\, \pi \, \omega\, L_1\, y} \, ,
\;\;\;\;\;\;\;\; U_2(x,y) = e^{-i\, \pi \, \omega\, L_2\, x}
\label{tf}
\ee
where
\be \omega = \frac{1}{\theta \pi} \left( 1 - \sqrt{ 1 + {2 \pi
\theta k}/{L_1 L_2}} \right) \; , \;\;\;\;\; k \in \mathbb{Z}
\label{tf1}\ee
It should be noted that in the $\theta \to 0$ limit, solution
(\ref{tf})-(\ref{tf1}) goes smoothly to the solution on the
commutative torus. One can make easily contact between this result
and the discussion in \cite{ambj}  on pure $U(p)$ Yang-Mills
theory on the noncommutative torus (in the particular $p=1$ and
zero 't Hooft twist case).

Since $U_1$ and $U_2$ are translation generators, then for any
arbitrary function $f(x,y)$ it holds that
\begin{align}
U_{1}(x,y)\, f(x,y) \, U^{-1}_1(x,y) = f(x+\pi \omega L_1
\theta, y) \nonumber \\
U_{2}(x,y)\, f(x,y) \, U^{-1}_2(x,y) = f(x, y +\pi \omega L_2
\theta) \label{translation}
\end{align}

Periodicity conditions (\ref{comm-1})  and the gauge transformation
laws  imply the following transformation laws for the
transition functions
under  gauge transformations
\begin{align}
U_1(x,y) \; &\to \; U'_1(x,y) = V(x+L_1,y)\, U_1(x,y)\,
V^{-1}(x,y)
\nonumber \\
U_2(x,y) \; &\to \; U'_2(x,y) = V(x,y+L_2)\, U_2(x,y)\,
V^{-1}(x,y)
\end{align}
   Now,
using property \eqref{translation} we have
\begin{align}
U'_1(x,y) &= V(x+L_1,y)\, V^{-1}(x+\pi \omega L_1 \theta, y)\,
U_1(x,y)
\nonumber \\
U'_2(x,y) &= V(x,y+L_2)\, V^{-1}(x, y +\pi \omega L_2 \theta) \,
U_2(x,y)
\end{align}
%
%%%%%%%%%%%%%%%%%%%%%%%%%%%%%%%%%%%%%%%%%%%%%%%%%%%%%%%%%%%%%%%%%%%%%%%%%%%
%%%%%%%%%%%%%%%%%%%%%%%%%%%%%%%%%%%%%%%%%%%%%%%%%%%%%%%%%%%%%%%%%%%%%%%%%%
%%%%%%%%%%%%%%%%%%%%%%%%%%%%%%%%%%%%%%%%%%%%%%%%%%%%%%%%%%%%%%%%%%%%%%%%%%%
%%%%%%%%%%%%%%%%%%%%%%%%%%%%%%%%%%%%%%%%%%%%%%%%%%%%%%%%%%%%%%%%%%%%%%%%%%
%
Then, if the gauge transformation functions are periodic with
periods
\be {\tilde L}_i = s\, L_i\; , \;\;\;\;\;\;\; s=(1 - \pi \omega
\theta) = \, \sqrt{1 + \frac{2 \pi \theta k}{L_1 L_2}} \,
\;\;\;\;\;\;\;\; i=1,2\label{ltilde0} \ee
the transition functions are invariant. Thus, we will restrict
ourselves to gauge transformations satisfying this property. From
now on, we shall call $\mathcal{\tilde T}$ the scaled torus with
periods $({\tilde L}_1, {\tilde L}_2)$.
%
% FIN AGREGADO
%%%%%%%%%%%%%%%%%%%%%%%%%%%%%%%%%%%%%%%%%%%%%%%%%%%%%%%%%%%%%%%%%%%%%%%%%%%
%%%%%%%%%%%%%%%%%%%%%%%%%%%%%%%%%%%%%%%%%%%%%%%%%%%%%%%%%%%%%%%%%%%%%%%%%%
%%%%%%%%%%%%%%%%%%%%%%%%%%%%%%%%%%%%%%%%%%%%%%%%%%%%%%%%%%%%%%%%%%%%%%%%%%%
%%%%%%%%%%%%%%%%%%%%%%%%%%%%%%%%%%%%%%%%%%%%%%%%%%%%%%%%%%%%%%%%%%%%%%%%%%

%
%%%%%%%%%%%%%%%%%%%%%%%%%%%%%%%%%%%%%%%%%%%%%%%%%%%%%%%%%%%%%%%%%%%%%%%%%%%
%%%%%%%%%%%%%%%%%%%%%%%%%%%%%%%%%%%%%%%%%%%%%%%%%%%%%%%%%%%%%%%%%%%%%%%%%%
%%%%%%%%%%%%%%%%%%%%%%%%%%%%%%%%%%%%%%%%%%%%%%%%%%%%%%%%%%%%%%%%%%%%%%%%%%
%%%%%%%%%%%%%%%%%%%%%%%%%%%%%%%%%%%%%%%%%%%%%%%%%%%%%%%%%%%%%%%%%%%%%%%%%%
%
%%%%%%%%%%%%%%%%%%%%%%%%%%%%%%%%%%%%%%%%%%%%%%%%%%%%%%%%%%%%%%%%%%%%%%%%%%%
%%%%%%%%%%%%%%%%%%%%%%%%%%%%%%%%%%%%%%%%%%%%%%%%%%%%%%%%%%%%%%%%%%%%%%%%%%
%%%%%%%%%%%%%%%%%%%%%%%%%%%%%%%%%%%%%%%%%%%%%%%%%%%%%%%%%%%%%%%%%%%%%%%%%%%
%%%%%%%%%%%%%%%%%%%%%%%%%%%%%%%%%%%%%%%%%%%%%%%%%%%%%%%%%%%%%%%%%%%%%%%%%%

The boundary conditions (\ref{bca}) together with our choice of
transitions functions (\ref{tf}) imply for the gauge field the
following equations
\begin{align}
A_{1}(x+L_1(1 - \pi \omega \theta) ,y) &=   A_{1}(x,y) \nonumber\\
A_{1}(x, y +L_2(1 - \pi \omega \theta)) &=   A_{1}(x,y) -
\frac{1}{g}\pi \omega L_2 \nonumber\\
A_{2}(x+L_1(1 - \pi \omega \theta) ,y) &=   A_{2}(x,y) +
\frac{1}{g}\pi \omega L_1 \nonumber\\
A_{3}(x, y +L_2(1 - \pi \omega \theta)) &=   A_{2}(x,y)%
\end{align}
which have as a general solution,
\be A_i(x,y) = {\tilde A}_i(x,y) + a_i(x,y) \label{generalA}\ee
where ${\tilde A}_i$ is a periodic function in the scaled torus
${\mathcal{\tilde T}}$
and  $a_i$ is defined as
\be a_i = f\, \varepsilon_{ij} x^j \ee
with
\be f = \frac{1}{g\theta} \left( 1-\frac{1}{s} \right)
\label{f}\ee
The field strength $F_{ij}=\partial_{i} A_{j} - \partial_{j} A_{i}
- i g\, [A_{i},A_{j}]\,$ can be written more conveniently as
\be F_{ij}= \frac{1}{s} {\tilde F}_{ij} + f_{ij}
\label{generalAB}\ee
where
\begin{align}
f_{ij}
&= -\varepsilon_{ij}\, \frac{2 \pi k}{g}\,\frac{1}{ {\tilde L}_1
{\tilde L}_2}
\end{align}
and
\begin{align}
{\tilde F}_{ij}
& = \partial_i {\tilde A}_j - \partial_j
{\tilde A}_i - i \, \tilde{g} \, [{\tilde A}_i,
{\tilde A}_j]
\end{align}
where we have introduced the scaled charge,

\be {\tilde g} = s\, g \ee and
 we have used that
\be \varepsilon_{ij}\, [x_j, ~] = -i \theta \, \partial_i
\label{deriv-ident}\ee
Let see how does the field $ A$   transforms under gauge
transformations. Applying a gauge transformation to
\eqref{generalA} we have
\begin{align} A'_i &= V\left({\tilde A}_i + a_i \right)V +
\frac{i}{g}\, V\, \partial_{i} \, V^{-1} \nonumber \\
&=V\,{\tilde A}_i \,V + g\,f\,\varepsilon_{ij}\, V\,x^j \, V^{-1}
+  \frac{i}{g}\, V\, \partial_{i} \, V^{-1}
\end{align}
But using \eqref{deriv-ident} we can rewrite the middle term as a
derivative term plus $a_i$
\begin{align}
A'_i &= V\,{\tilde A}_i \,V - i\, \theta f\,  V\,
\partial_{i} \, V^{-1} + a_i +
\frac{i}{g}\, V\, \partial_{i} \, V^{-1} \nonumber \\
&=V\,{\tilde A}_i \,V + i \frac{1-g\, \theta f}{g}\,  V\,
\partial_{i} \, V^{-1} + a_i \nonumber \\
&=V\,{\tilde A}_i \,V + \frac{i}{{\tilde g}}\,  V\,
\partial_{i} \, V^{-1} + a_i
\end{align}
Thus a gauge transformation on $A_i$ is equivalent to a gauge
transformation on ${\tilde A}_i$ but with the scaled charge
${\tilde g}$ (and the field $a_i$  untransformed).

We can summarize these results by stating  that a gauge theory on
the noncommutative torus $\mathcal{T}$ and with non-trivial
boundary conditions \eqref{bca} is equivalent to a gauge theory on
the scaled noncommutative torus $\mathcal{\tilde T}$, with
periodic boundary conditions and with a scaled charge ${\tilde
g}$.

Let us now solve the boundary condition equations for the Higgs
field. A field $\phi(x)$ satisfying the boundary conditions
\eqref{comm-1} with the transition functions given in equation
\eqref{tf}, can be decomposed as
\be \phi(x,y) = \phi_0(x,y)\, \eta(x,y) \label{phi-decomp}\ee
where $\phi_0(x,y)$ is an arbitrary function periodic in the
scaled torus $\mathcal{\tilde T}$ and $\eta(x,y)$ satisfy the same
boundary conditions as $\phi(x,y)$. Then we just have to find a
particular solution of
\begin{align}
\eta(x+L_1,y) &= U_1(x,y)\, \eta(x,y) \nonumber \\%
\eta(x,y+L_2) &= U_2(x,y)\, \eta(x,y) \label{app2-bceta}
\end{align}
Inspired in the commutative case \cite{GAR} let us consider a
function $h(x,y)$ of the form
\be h(x,y) = e^{i \alpha \left\{z\, , \, y\right\} } \;, \;
\;\;\;\;\;\;\;\;\; z = x + i y \label{app2-h1}\ee
where $\left\{z\, , \, y\right\} = z\, y + y\, z$ and $\alpha$ is
determined by the condition
\begin{align}
h(x+L_1,y) &= U_1(x,y)\, h(x,y) \nonumber \\
&=e^{i\, \pi \, \omega\, L_1\, y} \, e^{i \alpha \left\{z\, , \,
y\right\}} \label{sol-h}
\end{align}
Since $[z,y]= i \, \theta$, we can use the result
\eqref{haus-result1} of the appendix to obtain
\be U_1(x,y)\, h(x,y) = e^{i \alpha \left\{z+c\, , \, y\right\} }
\;\;, \;\;\;\;\;\;\;\;\;\; c =\frac{\theta \pi \omega L_1}{1-e^{-2
\theta \alpha}} \ee
Then, equation \eqref{sol-h} is solved if we chose
\be \alpha= -\frac{1}{2 \theta} \log\left(1 - \pi \omega
\theta\right) = -\frac{1}{2 \theta} \log s \ee
Now we compute
\begin{align}
U_2(x,y)\, h(x,y) =e^{-i\, \pi \, \omega\, L_2\,x} \, e^{i \alpha
\left\{z\, , \, y\right\}}
\end{align}
Using several times equations \eqref{haus-result1},
\eqref{haus-result2}, \eqref{haus-result3} and
\eqref{haus-result4} of the appendix we get
\begin{align}
U_2(x,y)\, h(x,y) &= e^{-k \pi L_2/L_1}\, e^{i \alpha \left\{z+i
L_2\, , \, y - {\tilde L}_2\right\} } \nonumber \\
&=e^{k \pi L_2/L_1}\, e^{i \alpha \left\{z+i L_2\, , \, y -
{L}_2\right\} } \, e^{-i 2\pi k \, z/L_1} \nonumber \\
&=h(x,y+L_2)\,e^{k \pi L_2/L_1}\, e^{-i 2\pi k \, z/L_1}
\end{align}
(we used that $\omega\left(1-\pi \omega \theta/2\right) = k/L_1
L_2$).  Then, $\eta(x,y)$ can be written as
\be \eta(x,y)=h(x,y) \, \Theta(x,y) \ee
with $\Theta(x,y)$ satisfying
\begin{align}
\Theta(x+L_1,y) &= \Theta(x,y) \nonumber \\
\Theta(x,y+L_2) &= e^{k \pi L_2/L_1}\, e^{-i 2\pi k \, z/L_1}\,
\Theta(x,y) \label{app2-bcTheta}
\end{align}
In commutative space, a function that satisfies
\eqref{app2-bcTheta} is given by a product of Riemann $\theta_3$
functions
\be \Theta(x,y) = \prod_{n=1}^k \theta_3\left(\pi (z + a_n)/L_1|i
L_1/L_2\right) \label{app2-Theta}\ee
where
\be \theta_3(z|\tau) = \sum_n e^{i\pi \tau n^2 + 2 i n z}
\label{theta-3}\ee
and $a_n$ are arbitrary complex numbers satisfying
\be \sum_{n=1}^k a_n = 0 \ee
(the function $\Theta(x,y)$ has $k$ zeros at the points
$a_i+(L_1+i L_2)/4$).

However, since the theta functions are only functions of one
variable, we can replace the standard product with the
noncommutative product, as they both coincide. Then, eq.
\eqref{app2-Theta} {\it is} the solution of \eqref{app2-bcTheta}
in noncommutative space. Thus,
\be \eta(x,y) = e^{i \alpha \left\{z\, , \, y\right\} } \,
\prod_{n=1}^k \theta_3\left(\pi (z + a_n)/L_1 |i L_1/L_2\right)
\label{eta-k} \ee
with
\be \alpha= -\frac{1}{2 \theta} \log s \ee
In the limit $\theta \to 0$ this function coincides with the one
obtained in the commutative case (see \cite{GAR}). For the special
case of $k=1$ we have
\be \eta(x,y) = e^{i \alpha \left\{z\, , \, y\right\} } \,
\theta_3\left(\pi z /L_1 |i L_1/L_2\right) \ee
%

%%%%%%%%%%%%%%%%%%%%%%%%%%%%%%%%%%%%%%%%%%%%%%%%%%%%%%%%%%%%%%%%%%%%%
%%%%%%%%%%%%%%%%%%%%%%%%%%%%%%%%%%%%%%%%%%%%%%%%%%%%%%%%%%%%%%%%%%%%%

In order to discuss the dynamics   through the introduction of the
action and the energy of our model, we have to define an
appropriate trace (or integral) on the noncommutative torus.
Calling  ${\cal A}_\theta$ the space of  functions defined on
${\cal T}$,  a generic  periodic function  $f(x,y)$ can be written
in the form
\be f(x,y) = \sum_{m,n} f_{mn} \exp\left(i m\frac{x}{L_1}\right)
\exp\left( i n\frac{y}{L_2} \right) \ee
and then one can formally define integration  in ${\cal
A}_\theta$, which we shall call trace ${\rm Tr}$, as follows
\be {\rm Tr} f(x,y) = f_{00}L_1L_2 \label{integral0} \ee
which in turns defines an integral over $\mathcal{T}$. This
operation satisfies ${\rm Tr} (fg) = {\rm Tr} (gf) $ and reduces
in the commutative limit to the standard integral on ${\cal T}$.

We have defined in \eqref{integral0} the integration in the
noncommutative torus of strictly periodic functions $f(x,y)$.
However the definition has to be corrected when the integrand
satisfies twisted boundary conditions \cite{Ho},\cite{BMZ}. We
discuss this issue in detail in Appendix 3 and here give a brief
summary. Consider a function $f(x,y)$ that satisfies twisted
boundary conditions in the adjoint section (as it is the case of
$F_{ij}$ for example)
\begin{align}
f(x+L_1,y) &=  U_{1}(x,y)\, f(x,y) \, U^{-1}_1(x,y)  \nonumber\\
f(x,y + L_2) &=  U_{2}(x,y)\,f(x,y) \, U^{-1}_2(x,y) %
\label{twisted-ad-bc}
\end{align}
Then, using \eqref{translation} we see that $f(x,y)$ is in fact
{\it periodic} in the scaled torus ${\mathcal{\tilde T}}$. So the
natural integration measure for the function $f(x,y)$ is on the
scaled torus ${\mathcal{\tilde T}}$, that is
\be I[f] = {\rm Tr}_ {\mathcal{\tilde T}} \, f \ee

It can be shown that this definition is crucial if we want to
preserve the cyclic property of the integral (trace) which is
essential in order to derive the equations of motion. Consider for
example two functions $\phi_1({\vec x})$ and $\phi_2({\vec x})$
that have nontrivial boundary conditions
\begin{align}
\phi_i(x+L_1,y) &= U_1(x,y)\, \phi_i(x,y) \nonumber \\%
\phi_i(x,y+L_2) &= U_2(x,y)\, \phi_i(x,y)\; , \;\;\;\;\;
\;\;\;\;\;\;\;\; i=1,2\label{phi1-2}
\end{align}
Then the product
\be \phi_1(\vec x)\, \phi_2^{\dagger}(\vec x) \ee
is strictly periodic in the torus $\mathcal{T}$, but the transpose
product,
\be \phi_2^{\dagger}(\vec x) \, \phi_1(\vec x) \ee
satisfy nontrivial boundary conditions in the adjoint, so it is
periodic in the scaled torus $\mathcal{\tilde T}$. Nonetheless, as
we show in the appendix, the cyclic property of the integral is
still valid provided we integrate the first function in
${\mathcal{T}}$ and the second one in ${\mathcal{\tilde T}}$
\be {\rm Tr}_ {\mathcal{ T}}  \left( \phi_1(\vec x)\,
\phi_2^{\dagger}(\vec x) \right) = {\rm Tr}_ {\mathcal{\tilde T}}
\left( \phi_2^{\dagger}(\vec x)\, \phi_1(\vec x)\right)
\label{cyclic-g}\ee
That is, the cyclic property is preserved with the above
definition.

%%%%%%%%%%%%%%%%%%%%%%%%%%%%%%%%%%%%%%%%%%%%%%%%%%%%%%%%%%%%%%%%%%%%%%%%%%%
%%%%%%%%%%%%%%%%%%%%%%%%%%%%%%%%%%%%%%%%%%%%%%%%%%%%%%%%%%%%%%%%%%%%%%%%%%
%%%%%%%%%%%%%%%%%%%%%%%%%%%%%%%%%%%%%%%%%%%%%%%%%%%%%%%%%%%%%%%%%%%%%%%%%%
%%%%%%%%%%%%%%%%%%%%%%%%%%%%%%%%%%%%%%%%%%%%%%%%%%%%%%%%%%%%%%%%%%%%%%%%%%
\section{The Maxwell-Higgs model}

We shall consider here a $U(1)$ gauge field coupled to a Higgs
scalar defined on the noncommutative torus. Dynamics of the model
is governed by the Lagrangian
\begin{equation}
L =   -\frac{1}{4}\,F_{\mu\nu} F^{\mu\nu} + (D_\mu
\Phi)^{\dagger}\, (D^\mu \Phi) - \lambda \, (\Phi^{\dagger}
\Phi-\phi_0^2)^2
\end{equation}
We are interested in static configurations so that   the energy
can be written in the form\footnote{In this expression we are
mixing covariantly periodic terms ($F_{ij} F_{ij}$) with strictly
periodic terms, ($(D_i \Phi)^{\dagger}\, (D_i \Phi)$ and
$(\Phi^{\dagger} \Phi-\phi_0^2)^2$), so according to the previous
discussion on integration, the integrals has to be defined in
their appropriate domains. Note however, that we can convert the
periodic terms into covariantly periodic ones by using property
\eqref{cyclic-g}, and thus, the whole Lagrangian or energy  have
to be integrated in the same domain, the scaled torus
$\mathcal{\tilde T}$.}
 \be E = {\rm Tr}  \,  \left( \frac{1}{4}\,F_{ij}
F_{ij} + (D_i \Phi)^{\dagger}\, (D_i \Phi) + \lambda \,
(\Phi^{\dagger} \Phi-\phi_0^2)^2\right) \ee
Here $D_i \Phi = \partial_i \Phi - i g A_i\, \Phi$ is the
covariant derivative and $F_{ij}$ is the electromagnetic tensor.
Notice that, via the covariant derivative, we are choosing for
definiteness a  Higgs-gauge coupling which corresponds to the
fundamental representation (other choices are possible).

As in the commutative case, the energy can be rewritten using the
Bogomolny trick as,
\begin{align}
E = & {\rm Tr}   \left( \frac{1}{2}\, |D_i \Phi - i \gamma \,
\varepsilon_{ij}\, D_j\Phi|^2 +  \frac{1}{4}\,\left( F_{ij} -
\gamma\, g\, \varepsilon_{ij} (\Phi\, \Phi^{\dagger}  - \,
\phi_0^2) \right)^2 + \right. \nonumber \\
&\left. \left(\lambda - \frac{g^2}{2} \right) \, \left(
\Phi^{\dagger}\, \Phi - \phi_0\right)^2 -  \gamma \, \frac{g}{2}\,
\phi_0^2 \,
\varepsilon_{ij} \, F_{ij} + \text{total derivative}\right ) \nonumber \\
\label{ener2}
\end{align}
where $\gamma=\pm 1$

The BPS equation corresponding to a bound of the energy when
$\lambda = g^2/2$, \be E \geq -\gamma \frac{g}{2} \phi_0^2 {\rm
Tr}_ {\mathcal{ T}} \,  \varepsilon_{ij}F_{ij} \ee
then read,
\begin{align}
& D_i \Phi - i \gamma \, \varepsilon_{ij}\, D_j\Phi=0
\\
& F_{ij} - \gamma\, g\, \varepsilon_{ij} (\Phi\, \Phi^{\dagger}  -
\, \phi_0^2)=0
\end{align}
Setting for definiteness  $\gamma=-1$ and using
(\ref{generalA})-(\ref{generalAB}), we can write the BPS equations
as
\be {\tilde F}_{12} = \, {\tilde g} \left( \Phi \Phi^{\dagger} -
\left(\phi_0^2 - \frac{2 \pi k}{ g^2 {\tilde L}_1 {\tilde
L}_2}\right) \right) \label{bo1} \ee
\be {\tilde D}_{\bar z}\Phi + \frac{\pi \, \omega}{2} \Phi \, z
=0
\label{bo2}
\ee
where $z=x+i y$.

Since the fields $\tilde A$ are periodic in the scaled torus
$\mathcal{\tilde T}$, the total flux of ${\tilde  F}_{ij}$ on
$\mathcal{\tilde T}$ vanishes (see equation \eqref{integral0}) and
then  we have
\begin{align}
\Phi &={\rm Tr}_ {\mathcal{\tilde T}} \,  F_{12} = {\rm Tr}_
{\mathcal{\tilde T}} \,  f_{12} = - \frac{2 \pi k}{g}
\end{align}

Bogomolny equations (\ref{bo1})-(\ref{bo2}) have the particular
solution
\be {\tilde A}=\Phi=0 \label{brado} \ee
provided the area of the torus and the Higgs vev are related
according to
\be \phi_0^2 = \frac{2 \pi k}{ g^2 {\tilde L}_1 {\tilde L}_2}
\label{brad}\ee
In the $\theta \to 0 $  commutative limit this solution reproduces
the so called Bradlow solution \cite{Bradlow} on the torus.
Moreover, as in the commutative case \cite{GAR},  solution
(\ref{brado})-(\ref{brad}) could then be used as a starting point
to obtain  new solutions with non-vanishing $\tilde A$ and $\Phi$,
by an appropriate expansion.

In order to search for general solutions to
eqs.(\ref{bo1})-(\ref{bo2}) it will be convenient to parametrize
the fields as
\be {\tilde A}_{\bar z} = \frac{i}{\tilde g} M^{-1} \partial_{\bar
z} M + {\tilde A}^0_{\bar z} \label{paramet-a}\ee
\be \Phi = M^{-1} \chi  \label{paramet-phi}\ee
where $M$ is a complex (non unitary) function periodic in
$\mathcal{\tilde T}$, ${\tilde A}^0_{\bar z}$ is a constant field,
and $\chi$ has the same periodicity as $\Phi$. The BPS equation
(\ref{bo2}) then becomes,
\be \partial_{\bar z} \chi - i {\tilde g} {\tilde A}^0_{\bar z}
\chi + \frac{\pi \, \omega}{2} \chi \, z = 0 \label{bo3}\ee
As we showed previously in equation \eqref{phi-decomp}, the
function $\chi$ can be factorized as
\be \chi(x,y) = \chi_0(x,y)\, \eta(x,y) \label{factor-phi}\ee
where $\eta$ carries the non trivial boundary conditions (see eq.
\eqref{eta-k}) and $\chi_0$ is periodic in $\mathcal{\tilde T}$.
Replacing \eqref{factor-phi} in \eqref{bo3} we get
\be \left(\partial_{\bar z} \chi_0 - i {\tilde g} {\tilde
A}^0_{\bar z} \chi_0\right) \eta  + \chi_0 \left( \partial_{\bar
z} \eta + \frac{\pi \, \omega}{2} \, z \right)= 0 \label{bo4}\ee
To compute $\partial_{\bar z} \eta$ we first use equations
\eqref{haus-result3} and \eqref{haus-result4} of the appendix to
rewrite
\be \eta(x,y) = e^{-\alpha \{z,{\bar z}\}} \, e^{ \pi k z^2/2 L_1
L_2} \, \Theta(z) \ee
where $\Theta$, given in equation \eqref{app2-Theta}, in only
function of $z$. Thus the problem reduces to compute the
derivative with respect to $\bar z$ of $ e^{-\alpha \{z,{\bar
z}\}}$. Using that
\be \partial_{\bar z} = \frac{1}{2 \theta}\, [z,\,] \ee
and that $[z, {\bar z}]= 2 \theta$  we can show that
\be \partial_{\bar z}  e^{-\alpha \{z,{\bar z}\}} = -\frac{\pi
\omega}{2}\, e^{-\alpha \{z,{\bar z}\}} \, z \ee
and then the second term of equation \eqref{bo4} vanishes. So, the
BPS equation (\ref{bo2}) reduces to
\be \partial_{\bar z} \chi_0 - i {\tilde g} {\tilde A}^0_{\bar z}
\chi_0 = 0 \ee
with solution
\be \chi_0 = N\, e^{i {\tilde g} \left(i {\tilde A}^0_{\bar z}
{\bar z} + {\tilde A}^0_{z} z \right)} \ee
where $N$ is a normalization factor. Periodicity of $\chi_0$
requires that ${\tilde A}^0$ has the form
\be {\tilde A}^0_{z}=\frac{\pi}{\tilde g} \left(
\frac{n_0}{{\tilde L}_1} + i \frac{m_0}{{\tilde L}_2} \right)
\label{def-a0}\ee
with integers $n_0$ y $m_0$. In commutative space this particular
form of $A_0$ is a pure gauge and thus can be simply gauged away.
In noncommutative space this is also the case with the proviso
that the gauge transformation will also transform the non-trivial
part of the field $\tilde A$ (equation \eqref{paramet-a}). However
the effect of the transformation will be only a shift in the
coordinates of the fields. So, without losing generality we can
make $m_0=n_0=0$.
\\

%%%%%%%%%%%%%%%%%%%%%%%%%%%%%%%%%%%%%%%%%%%%%%%%%%%%%%%%%%%%%%%%%%%%%%%%%%%%%%

Concerning the BPS equation eq.(\ref{bo2}), one has first to write
the field ${\tilde {\tilde F}}_{12}$ in terms of the variables $M$
defined in eq. (\ref{paramet-a}). Clearly the gauge invariant
variables are to be  defined from the combination
\be H=M\, M^{\dagger} \ee
so that one should be able to write the Bogomolny equations in
terms of $H$.  Since ${ {\tilde F}}_{12}$ is not gauge invariant
but covariant, one can not write it only in terms of $H$; indeed a
straightforward computation gives
\be {\tilde F}_{z {\bar z}}  = \frac{i}{\tilde g} M ^{-1}\,H\,
\partial_{z}\left(H^{-1}\partial_{\bar z} H \right) M^{\dagger {-1}}\ee
Substituting this expression, and that for  $\Phi$ given by
eq.(\ref{paramet-phi}) in the Bogomolny equation leads to
\be H\,
\partial_{z}\left(H^{-1}\partial_{\bar z} H \right)=
 \frac{1}{2} \,{\tilde g}^2 \left( \chi\, \chi^{\dagger} - \mu_0^2
H\right) \label{chose}\ee
where $\chi$ is given in equations
(\ref{factor-phi})-(\ref{def-a0}) and
\be \mu_0^2 = \phi_0^2 - \frac{2 \pi k}{ g^2 {\tilde L}_1 {\tilde
L}_2}
\ee
In order to make further progress to find solutions of
eq.(\ref{chose}) one has in principle to resort to numerical
techniques as it is already the cased for $\theta = 0$.

\section{Non Abelian extension and discussion}
It should be possible to extend most of our results  to the case
of (appropriate) non Abelian gauge groups. As it is well known,
consistency of noncommutative theories requires to work with
$U(N)$ groups and not $SU(N)$ \cite{Chaichian}. One can then
consider a $U(2)\times U(1)$ model as a first step in the study of
vortex solutions in a noncommutative version of the standard
model, along the lines of Ref.\cite{BL1} for the commutative case.

Consider  then the energy for static configurations,
\be E = {\rm Tr}_ {\mathcal{ T}}  \,  \left( \frac{1}{2}\,{\rm tr}
(W_{ij}\,W_{ij}) + \frac{1}{4}\, tr (B_{ij}B_{ij} )   + (D_i
\Phi)^{\dagger}\, (D_i \Phi) + \lambda \, (\Phi^{\dagger}
\Phi-\phi_0^2)^2\right) \ee
where the $U(2)$ gauge fields are defined as
\be W_i = W_i^a\, \lambda^a \,  , \;\;\;\;\lambda^0 =
\frac{1}{2}\,  I \, , \;\; \lambda^k = \frac{1}{2}\, \sigma^k \ee
$B_i$ is a $U(1)$ gauge field, $\Phi$ is a Higgs field in the
fundamental representation of $U(2)$ and the covariant derivatives
and field strengths are defined as
\be D_i \Phi = \partial_i \Phi - i g W_i\, \Phi + i \frac{g'}{2}
\Phi \, B_i \ee \be W_{ij}= \partial_i W_j - \partial_j W_i + ig
[W_i,W_j]\, ,  \,\,\,\,\,\,
 B_{ij}= \partial_i B_j - \partial_jB_i +ig'[B_i,B_j]
\ee Notice that the covariant derivative is defined so that it
acts from the left for the $U(2)$ group and from the right for the
$U(1)$ one. The appropriate way to write perfect a  square  \`a la
Bogomolny for the Higgs covariant derivative is in this case
\begin{eqnarray}
 |D_i\Phi|^2& =& |D_i \Phi - i \gamma \, \varepsilon_{ij}\,
D_j\Phi|^2 - \gamma \,g  \, {\rm tr} (\varepsilon_{ij}
\Phi^{\dagger}\, W_{ij}\, \Phi ) + \gamma \frac{g'}{2}
(\Phi^{\dagger} \Phi)\,
\varepsilon_{ij}\, B_{ij} + \nonumber\\
&~& +   {\rm ~divergence}
 \end{eqnarray}
leading to the following expression for the energy:
\begin{align}
E = &{\rm Tr}  \,   \, \left( \frac{1}{2}\, |D_i \Phi - i \gamma
\, \varepsilon_{ij}\, D_j\Phi|^2 + \frac{1}{2}\, {\rm tr} \left(
W_{ij} - \gamma \varepsilon \, \frac{g}{2} \Phi\, \Phi^{\dagger}
\right)^2 + \right.
\nonumber \\
&  \frac{1}{4}\,\left( B_{ij} + \gamma\, \frac{g'}{2}\,
\varepsilon_{ij} (\Phi^{\dagger} \Phi - \mu^2 \, \phi_0^2)
\right)^2 + \gamma \, \frac{g'}{2}\, \mu^2 \phi_0^2 \,
\varepsilon_{ij} \, B_{ij} + \nonumber \\
& \left(\lambda - \frac{g^2}{4} - \frac{g'^2}{8}\right) \, \left(
\Phi^{\dagger}\, \Phi - \phi_0\right)^2 - \left( \frac{g^2}{2} +
\frac{g'^2}{4}(1 - \mu^2) \right)  \Phi^{\dagger}\, \Phi\,
\phi_0^2 +  \nonumber \\
&\left. \left( \frac{g^2}{2} + \frac{g'^2}{4}(1 - \mu^4) \right)
\phi_0^4 \right)
\end{align}
Then, if we choose \be \mu^2=1+2\frac{g^2}{g'^2} \,\,\,\,\,
\lambda = \frac{g^2}{4} + \frac{g'^2}{8} \ee the energy is bounded
as \be E \geq \gamma \, g'\, \mu^2 \phi_0^2 \Phi_B -
 \mu^2( \mu^2-1)
\phi_0^4 A \ee where $\Phi_B$ is the flux of the $B$ field and $A$
is the area of the torus.

The bound is attained when the following  BPS equations are
satisfied,
\begin{align}
& D_i \Phi - i \gamma \,
\varepsilon_{ij}\, D_j\Phi=0 \\
&B_{ij} + \gamma\, \frac{g'}{2}\,
\varepsilon_{ij} (\Phi^{\dagger} \Phi - \mu^2 \, \phi_0^2)=0 \\
&W_{ij} - \gamma \varepsilon_{ij} \, \frac{g}{2} \Phi\, \Phi^{\dagger}
=0
\end{align}
As in the commutative space case \cite{BL1}-\cite{GAR}, the bound
has  a topological component, proportional to the $B$ flux and a
geometrical part, proportional to the area of the torus. The
non-commutative nature of space and the extra $U(1)$ factor
associated to the $U(2)$ group renders nevertheless, the analysis
of the solutions of these equations considerably more involved.

~

Let us end our work by summarizing our main results. We have
analyzed periodic configurations of matter and gauge fields in non
commutative space. We have discussed in detail how as a result of
coordinate non commutativity, the region of periodicity of gauge
invariant and gauge covariant quantities may differ, a property
that has to be kept in mind in order to obtain consistent results.
In this work, we have focussed mainly in the Abelian Maxwell Higgs
model, where we have been able to obtain BPS equations whose
vortex solutions also solve the Euler Lagrange equations. We have
presented a particular solution to these equations which, in the
$\theta \to 0 $ commutative limit corresponds the  Bradlow
solution on the commutative torus. In the general case, we were
able to reduce the problem of the two coupled BPS equation to that
of equation (\ref{chose}), which in principle should be  solved
using numerical techniques, as it is the case for the commutative
torus \cite{GAR}.

We believe that the generalization to non-Abelian models will not
present major difficulties. As a particular example and as a first
step in this direction, we have shown how the BPS equations of a
$U(2)\times U(1)$, a simplified version of the Standard Model in
non commutative space, are obtained. Of course, the noncommutative
character both of the space and the gauge group makes the
obtention of explicit solutions much more complicated but a more
detailed analysis should reveal the existence of Z-vortex arrays
(possibly with the presence of charged mesons condensates) as it
is the case in ordinary space \cite{BL1}. We hope to report on
this issues in a future publication.

\vspace{1 cm}

\noindent\underline{Acknowledgements}: This work  was partially
supported by UNLP, CICBA, CONICET, ANPCYT (PICT grant 03-05179)
Argentina and ECOS-Sud Argentina-France collaboration (grant
A01E02). F.A.S wishes to thank the Marie de Paris for the support
of a Senior Fellowship during part of this work.
\newpage

 %%%%%%%%%%%%%%%%%%%%%%%%%%%%%%%%%%%%%

\section*{Appendix}

\noindent {\bf A useful result}

\vspace{0.15 cm}

 \noindent We prove here a helpful result that was
used extensively throughout the paper:

~

\noindent{Lemma:}
{\sl Let $A$ and $B$ be two operators such that $[A,B]=i\mu A$
where $\mu$ is an arbitrary constant, then}
\be e^{i A}\,e^{i B} = e^{i \left(f(\mu) A + B\right)}
\label{haus-result1}\ee
{\sl where}
\be f(\mu) = \frac{\mu}{e^{\mu}-1} \ee

~

\noindent Proof: We write
\be e^{i A}\, e^{i B} = e^{i C} \label{haus-C}\ee
First we notice that a quick look at the Campbell-Baker-Hausdorff
formula
\be e^{A} e^{B} = e^{A + B + \frac{1}{2}[A,B] +
     \frac{1}{12}([A,[A,B]] + [B,[B,A]]) + \cdots} \ee
reveals that $C$ must be of the form
\be C= f(\mu) A + B \label{haus-C2}\ee
since any arbitrary nested commutator with $[A,B]$ will give,
either zero or something proportional to $A$. So the problem
reduces to find the function $f(\mu)$.

\nn Consider now the function
\be U(s) = e^{i s A}\, e^{i s B} \label{haus-Us}\ee
we have that
\be  \frac{d U}{ds} U^{-1} =  i \left(A + U\, B\,U^{-1}\right)
\ee
But
\begin{align}
 e^{i s A}\, B\, e^{-i s A} &= B + i s[A,B] \nonumber \\
 &= B - s \mu A
\end{align}
since higher order commutators vanish. Thus we have
\be  \frac{d U}{ds}=i\left( \left(1 - s\, \mu \right) A + B
\right) \, U \label{haus-derU0}\ee

Now we write according to \eqref{haus-C} and \eqref{haus-C2}
\be U(s) = e^{i C(s)}\; , \;\;\;\;\;\;\;\; C(s) = s\left( f(s \mu)
A + B\right) \ee
We have
\be \frac{d U}{ds} = \sum_{n=0}^{\infty} \frac{i^n}{n!}\, \frac{d
C(s)^n}{ds} \label{haus-derU}\ee
but
\begin{align}
\frac{d C(s)^n}{ds} &= \sum_{p=0}^{n-1} C^p\,\frac{d C(s)}{ds} \,
C^{n-p} \nonumber \\
&=s^{-1}\,n\,C(s) +  \mu f'(s \mu)\,\sum_{p=0}^{n-1} C^p\, A \,
C^{n-p} \label{haus-derC}
\end{align}
Now we notice that
\begin{align}
C(s)\, A &= A\, C(s) + [C(s),A] \nonumber \\
&= A\,\left(C(s) - i\,s\, \mu \right)
\end{align}
and applying successively this result we have
\be C(s)^p\,A = A\, \left(C(s) - i\,s\, \mu \right)^p \ee
Replacing this result back in \eqref{haus-derC} and then in
\eqref{haus-derU} we get
\begin{align}
\frac{d U}{ds} =i\, s^{-1}\, C(s)\,U +  \mu f'(s \mu)\,A\,
\sum_{n=0}^{\infty}  \frac{i^n}{n!}\,\sum_{p=0}^{n-1} \left(C(s) -
i\,s\, \mu \right)^p C(s)^{n-p-1}
\label{haus-sum2}
\end{align}
The sum in $p$ is a geometric sum, so it can be easily performed.
It gives
\be \sum_{p=0}^{n-1} \left(C(s) - i\,s\, \mu \right)^p
C(s)^{n-p-1} = (i\,s\, \mu)^{-1}\, \left( C(s)^n - \left(C(s)-
i\,s\, \mu \right)^n\right) \ee
and substituting this result in \eqref{haus-sum2} we get
\begin{align}
\frac{d U}{ds} &=i\, s^{-1}\, C(s) \,U +  \mu f'(s \mu)\,A\,
\sum_{n=0}^{\infty}  \frac{i^n}{n!}\,(i\,s\, \mu)^{-1}\, \left(
C(s)^n - \left(C(s)- i\,s\, \mu \right)^n\right) \nonumber \\
&=i\, s^{-1}\, C(s) \,U -i\, s^{-1} f'(s \mu)\,A
\left(e^{iC(s)} - e^{i(C(s) - i s \mu)}\right) \nonumber \\
&= i\left( \left(f - f' \left(1-e^{s \mu}\right) \right)A + B
\right)\, U
\label{haus-sum3}
\end{align}
Finally, comparing this equation with \eqref{haus-derU0} we have
the following differential equation for $f$
\be f - f' \left(1-e^{s \mu}\right) = 1 - s\,\mu\ee
The solution (with the initial condition $f(0)=1$, as can be
deduced from the series expansion of \eqref{haus-Us}) is
\be f(s \mu) = \frac{s \mu}{e^{s \mu} -1} \ee
and evaluating in $s=1$ we get the desired result.

\nn Taking the inverse of expression \eqref{haus-result1} (and
rescaling the fields and $\mu$) we have the equivalent result:
\be e^{i B}\,e^{i A} = e^{i \left(g(\mu) A + B\right)} \;,
\;\;\;\;\;\;\;\;\;\;\; g(\mu) = \frac{{\mu}}{1-e^{-\mu}}
\label{haus-result2}\ee
Similarly we can prove,
\be
e^{i\left(A + B\right)} = e^{i h(\mu) A}\, e^{i B}
\; ,\;\;\;\;\;\;\; h(\mu)=\frac{e^{\mu}-1}{\mu}\label{haus-result3}\\
\ee
and
\be e^{i\left(A + B\right)}= e^{i B}\, e^{i k(\mu) A}\, \;
,\;\;\;\;\;\;\;\; k(\mu)=\frac{1-e^{-\mu}}{\mu}
\label{haus-result4}\ee
In all cases
\be [A,B]=i\mu A \ee
%

%%%%%%%%%%%%%%%%%%%%%%%%%%%%%%%%%%%%%%%%%%%%%%%%%%%%%%%%%%%%%%%%%%%%%%
%%%%%%%%%%%%%%%%%%%%%%%%%%%%%%%%%%%%%%%%%%%%%%%%%%%%%%%%%%%%%%%%%%%%%%
%%%%%%%%%%%%%%%%%%%%%%%%%%%%%%%%%%%%%%%%%%%%%%%%%%%%%%%%%%%%%%%%%%%%%%
%%%%%%%%%%%%%%%%%%%%%%%%%%%%%%%%%%%%%%%%%%%%%%%%%%%%%%%%%%%%%%%%%%%%%%

~

\noindent{\bf Cyclic property of the integral }

\vspace{0.15 cm}

\noindent We will show below that the cyclic property of the
integral is valid whenever one defines the integration on the
appropriate torus.

First, from the definition of the integration on the torus, it is
straightforward to see that for strictly periodic functions
$f(x,y)$ and $g(x,y)$, the cyclic property holds
\be \int_{\mathcal{T}} f\,g = \int_{\mathcal{T}} g\,f \ee

Consider now two functions $\phi_1(x,y)$ and $\phi_2(x,y)$
satisfying the non-trivial periodic conditions
\begin{align}
\phi_i(x+L_1,y) &= U_1(x,y)\, \phi_i(x,y) \nonumber \\
\phi_i(x,y+L_2) &= U_2(x,y)\, \phi_i(x,y) \;\; ,
\;\;\;\;\;\;\;\;\;\;\;\;\;\;i=1,2\label{app3-phibc}
\end{align}
The product
\be w_1(x,y) = \phi^{\dagger}_1(x,y) \, \phi_2(x,y) \ee
is periodic on the torus $\mathcal{T}$. On the other hand the
reversed product
\be w_2(x,y) = \phi_2(x,y) \, \phi^{\dagger}_1(x,y) \ee
is periodic in the scaled torus $\mathcal{\tilde T}$
\eqref{twisted-ad-bc}. Thus, following the definition of integral
on the noncommutative torus, $w_1$ must be integrated on
$\mathcal{T}$ and $w_2$ on $\mathcal{\tilde T}$. We will show that
this definition satisfies
\be \int_{\mathcal{T}}w_1(x,y) = \int_{\mathcal{\tilde T}}
w_2(x,y) \ee
Let us consider, for simplicity, the case $k=1$. As we showed
previously, the functions $\phi_1$ and $\phi_2$ can be decomposed
as
\begin{align}
\phi_1(x,y) &= \phi^0_1(x,y)\, \eta(x,y) \nonumber \\
\phi_2(x,y) &= \phi^0_2(x,y)\, \eta(x,y)
\end{align}
where $\phi^0_i(x,y)\;, \;i=1,2$ are periodic in $\mathcal{\tilde
T}$ and
\be \eta(x,y) = e^{i \alpha \left\{z\, , \, y\right\} } \,
\theta_3\left(\pi z /L_1 |i L_1/L_2\right) \;,
\;\;\;\;\;\;\;\;\;\; \alpha= -\frac{1}{2 \theta} \log s\ee
Then
\begin{align}
w_1(x,y)=\eta^{\dagger}(x,y)\,\phi^{0\, \dagger}_1(x,y) \,
\phi^0_2(x,y)\, \eta(x,y) \nonumber \\
w_2(x,y)=\phi^0_2(x,y)\, \eta(x,y) \eta^{\dagger}(x,y)\,\phi^{0\,
\dagger}_1(x,y)
\end{align}
Consider first the integral
\be I_1 = \int_{\mathcal{T}} w_1(x,y) = \int_{\mathcal{T}}
\eta^{\dagger}(x,y)\,\phi^{0\, \dagger}_1(x,y) \, \phi^0_2(x,y)\,
\eta(x,y)\ee
Since $\,\phi^{0\, \dagger}_1(x,y) \, \phi^0_2(x,y)$ is periodic
in $\mathcal{\tilde T}$, without loss of generality we can replace
it by
\be e^{i 2\pi \left(n x/{\tilde L}_1 + m y/{\tilde L}_2\right)}\;
,\;\;\;\;n,m\in \mathbb{Z} \ee
Consider now the product
\be \gamma(x,y)=\eta^{\dagger}(x,y)\, e^{i 2\pi \left(n x/{\tilde
L}_1 + m y/{\tilde L}_2\right)} \eta(x,y)\ee
It can be easily shown that for any function $f(x,y)$
\be f(x,y)\,e^{i 2\pi \left(n x/{\tilde L}_1 + m y/{\tilde
L}_2\right)} = e^{i 2\pi \left(n x/{\tilde L}_1 + m y/{\tilde
L}_2\right)}\, f(x-t_1,y+t_2) \ee
where
\begin{align}
t_1 = 2 \pi m \,\theta/{\tilde L}_2 \; ,\;\;\;\; t_2 = 2 \pi \,n
\theta/{\tilde L}_1
\end{align}
Thus
\begin{align}
\gamma(x,y) &= e^{i 2\pi \left(n x/{\tilde L}_1 + m y/{\tilde
L}_2\right)} \, \eta^{\dagger}(x-t_1,y+t_2)\, \eta(x,y) \nonumber
\\
&= \sum_{p} \sum_{q} e^{i 2\pi \left(n x/{\tilde L}_1 + m
y/{\tilde L}_2\right)} \, e^{-\pi L_2/L_1 p^2 + i 2 \pi p
({\bar z}-t_1 - i t_2)/L_1} \times \nonumber \\
& \qquad  e^{-i \alpha \left\{{\bar z}-t_1 - i t_2\, , \,
y+t_2\right\}}\, e^{i \alpha \left\{z\, , \, y\right\}} \, e^{-\pi
L_2/L_1 q^2 + i 2 \pi q (z-t_1 + i t_2)/L_1}
\end{align}
Next we have to expand this expression in Fourier modes
\be \gamma(x,y) = \sum_{p,q} \gamma_{pq}\, e^{i 2\pi \left(p
x/{\tilde L}_1 + q y/{\tilde L}_2\right)} \ee
and keep the coefficient $\gamma_{00}$. Using several times the
identities \eqref{haus-result1}, \eqref{haus-result2},
\eqref{haus-result3}, and \eqref{haus-result4}, and after a
straightforward but long computation, we get
\be \gamma_{0 0} = \left\{
\begin{array}{ll}
0 & \text{if} \;\; m\; \text{is even}\\
\\
s\,\sqrt{2 L_1/L_2} e^{-\pi^2 \left(L_1^2 m^2 + L_2 n^2\right)/2
{\tilde L}_1 {\tilde L}_2 } \;\;\; & \text{if} \;\; m\; \text{is odd}\\
\end{array}
\right. \ee
Then
\be I_1 = \int_{\mathcal{T}} \gamma(x,y) = L_1 L_2\, \gamma_{0 0}
\ee

Now consider the integral
\be I_2 = \int_{\mathcal{\tilde T}} w_2(x,y) =
\int_{\mathcal{\tilde T}}  \phi^0_2(x,y)\, \eta(x,y)\,
\eta^{\dagger}(x,y)\, \phi^{0\, \dagger}_1(x,y)\ee
Since the product $\eta(x,y)\, \eta^{\dagger}(x,y)$ is periodic in
the torus $\mathcal{\tilde T}$, as well as $\phi^{0\,
\dagger}_1(x,y)$ and $\phi^{0}_2(x,y)$, we can rewrite $I_2$ as
\be I_2 = \int_{\mathcal{\tilde T}}  \phi^{0\, \dagger}_1(x,y)\,
\phi^0_2(x,y)\, \eta(x,y)\, \eta^{\dagger}(x,y) \ee
and again replace $\phi^{0\, \dagger}_1(x,y)\, \phi^0_2(x,y)$ by
$e^{i 2\pi \left(n x/{\tilde L}_1 + m y/{\tilde L}_2\right)}$.

First we compute
\begin{align}
\eta(x,y)\, \eta^{\dagger}(x,y) &= \sum_{p} \sum_{q}  e^{i \alpha
\left\{z\, , \, y\right\}}\, e^{-\pi
L_2/L_1 p^2 + i 2 \pi p z/L_1} \times \nonumber \\
& \hspace{-1cm}  \,  \, e^{-\pi L_2/L_1 q^2 + i 2 \pi q {\bar
z}/L_1} \, e^{-i \alpha \left\{{\bar z}\, , \, y\right\}}
\end{align}
After another long computation, using the identities
\eqref{haus-result1}, \eqref{haus-result2}, \eqref{haus-result3},
and \eqref{haus-result4}, we can write
\be \eta(x,y)\, \eta^{\dagger}(x,y) = s^{-1}\,\sqrt{2 L_1/L_2}\,
\sum_{p,q} e^{-\pi(L_2^2 p^2 + 4 L_1^2 q^2)/2 s^2}\, e^{i 2\pi
\left(p x/{\tilde L}_1 + 2 q y/{\tilde L}_2\right)}\ee
Thus, $\delta_{00}$, the $(0,0)$ Fourier mode of the product
\be \delta(x,y)= e^{i 2\pi \left(n x/{\tilde L}_1 + m y/{\tilde
L}_2\right)} \, \eta(x,y)\, \eta^{\dagger}(x,y)\ee
is given by
\be  \delta_{00} = \left\{
\begin{array}{ll}
0 & \text{if} \;\; m\; \text{is even}\\
\\
s^{-1}\,\sqrt{2 L_1/L_2} e^{-\pi^2 \left(L_1^2 m^2 + L_2
n^2\right)/2
{\tilde L}_1 {\tilde L}_2 } \;\;\; & \text{if} \;\; m\; \text{is odd}\\
\end{array}
\right. \ee
Notice that $\delta_{00}$ differs from $\gamma_{00}$ in a factor
$s^2$ which precisely the relation between the area of the two
torus.  The integral is
\be I_2 = \int_{\mathcal{\tilde T}}  w_2(x,y) = {\tilde L}_1
{\tilde L}_2\, \delta_{00} = L_1 L_2 \, s^2\,\delta_{00} \ee
So
\be \int_{\mathcal{T}}\phi^{\dagger}_1(x,y) \, \phi_2(x,y) =
\int_{\mathcal{\tilde T}} \phi_2(x,y) \, \phi^{\dagger}_1(x,y) \ee
%

%%%%%%%%%%%%%%%%%%%%%%%%%%%%%%%%%%%%%%%%%%%%%%%%%%%%%%%%%%%%%%%%%

\end{document}